\documentclass[aps,prd,twocolumn,superscriptaddress]{revtex4}
\usepackage{graphicx}
\usepackage{bm}
\usepackage{amsmath,mathrsfs}
\usepackage{amssymb}
\usepackage{slashed}
\usepackage{latexsym}
\usepackage{epsfig}
\usepackage{amsbsy}
\usepackage{array}
\usepackage{amssymb}
\usepackage{changes}
\usepackage{color}
\usepackage{setspace}
\usepackage{lipsum}
\usepackage{float}
\usepackage{color}
\usepackage{changes}
\usepackage[T1]{fontenc}
\usepackage{mathptmx}

\begin{document}

\title{  Exploring  hybrid   equation of state with constraints from  tidal deformability of GW170817 }

\author{Qing-Wu Wang }
 \email{qw.wang@scu.edu.cn }
 \affiliation{College of Physics, Sichuan University, Chengdu 610064, China}

  \author{Chao Shi}
  \email{ shichao0820@gmail.com}
  \affiliation{Department of nuclear science and technology,
Nanjing University of Aeronautics and Astronautics, Nanjing 210016, China;}

\author{Yan Yan}
\email{2919ywhhxh@163.com}
\affiliation{School of Mathematics and Physics, Changzhou University, Changzhou, Jiangsu 213164, China}

\author{Hong-Shi Zong}
\email{zonghs@nju.edu.cn}
\affiliation{Department of Physics, Nanjing University, Nanjing 210093, China}
\affiliation{Department of Physics, Anhui Normal University, Wuhu, Anhui 241000, China}
\affiliation{Nanjing Institute of Proton Source  Technology  , Nanjing 210046}

\begin{abstract}
With a   interpolation method on the P-$\mu$ plane, a   hybrid equation of state is explored.  The quark phase is described by our newly developed self-consistent two-flavor Nambu$-$Jona-Lasinio model. It  retains the contribution from the vector channel in the  Fierz-transformed Lagrangian  by introducing a weighting parameter  $\alpha$   [Chin. Phys. C \textbf{43}, 084102 (2019)]. In the hadron phase we use the relativistic mean-field   theory.    We   study the dependence of hybrid EOS and mass-radius relation   on $\alpha$.  It is found  that increasing $\alpha$ makes the hybrid EOS softer in the medium pressure. We can get stellar mass larger than $2M_\odot$. Further, we calculate the tidal deformability  $\tilde\Lambda$ for binary stars and compare with recent analysis GW170817 [Phys. Rev. X \textbf{9}, 011001  (2019)].


\end{abstract}

\maketitle
\section{Introduction} 

The phase transition of strongly interacting matter is an important topic in hadron physics. As the temperature and density   increase, the strongly interacting  matter will undergo a  phase transition  from hadronic phase to quark-gluon plasma (QGP), which is deconfined, approximate chiral symmetric state,   superfluid and  superconductivity, etc \cite{Buballa,Wilczek,Luo}.  It is generally believed that quarks exist in a hadronic state within a few times the saturation density of nuclear matter. However, how many times  is still an open question. This is due to our lack of limited density experimental data. From quark phase to hadron phase, the transition at high temperature and  low density may occur as a crossover \cite{Aoki2006N,Bali2012,Braun2011,Typel}. But  whether  there  is  a first-order phase transition at zero temperature and   high density or  not, along with the existence of  the critical  end point  are  still very unclear.
Many works indicate that a  quarkyonic zone ( chiral symmetry is partly restored but quark is still confined) may exist in the phase diagram \cite{Schaefer, Shao, Sakai}.  If so, it will affect chemical equilibrium at the hadron-quark phase transition. Unfortunately, the lattice simulation at present is not successful in exploring high chemical potential regions.   What's more serious is that experiments on Earth in the foreseeable future will have difficulty entering high density.

 The boom in  the astronomical observations of pulsars and neutron star mergers provides the possibility to check  theoretic models of hadron physics. Phase transition in neutron stars has links to gravitational waves \cite{Most,Orsaria,Hanauske}.
Astronomically, it is not easy to determine if phase transitions happened in the inner core of compact stars.   Massive stars with quark core are possibly less abundant in the first place.  If the observed pulsars are only pure neutron stars or quark stars, there is no possibility to observe any effect of hadron-quark phase transition.    In addition, the hadron-quark transition  could also be  a crossover at low temperature and high density.
 In Ref. \cite{Bauswein}, a peak of postmerger gravitational waves frequency ($f_{peak}$) is used to identify a first-order phase transition in the interior of neutron stars. But it  requires highly precise measurements of the masses and tidal deformabilities.

  The discovery of high-mass neutron stars has eliminated   a lot of models that provided soft equation of state (EOS) \cite{Demorest, Fonseca, Antoniadis, Cromartie,Abbott}.  Without modification in the Tolman-Oppenheimer-Volkoff (TOV) equations, only a sufficiently stiff equation of state can support high-mass neutron star.    It has been previously thought that stable quark matter should contain strange quarks \cite{Witten, Bodmer, Terazawa}, and  strange quark stars has become a hot issue.  Recent studies have shown that the two-flavor light quark matter can still exist stably \cite{Holdom,wangqy}.  On the other hand,  the introduction of strange hadrons will soften the EOS which makes it hard to obtain high-mass compact star \cite{Glendenning,Panda,Xu}. It is reasonable to construct a hybrid EOS  not containing strange quarks. In Ref. \cite{Montana},  the authors have found that only data from GW170817 \cite{Abbott2016} is compatible with the existence of hybrid stars.


Two methods are usually used in constructing the hadron-quark phase transition EOS \cite{Glendenningb,Yan,Benic}, i.e., the Maxwell construction and the Gibbs construction.  In the Maxwell construction  separate  neutrality of electric charge is required for the phase transition, but the energy and baryon densities can take different values.
 It  always results in a first-order phase transition  and may lead to mass-twin stars \cite{ Gerlach,Kaltenborn,Blaschke}. In contrast, for the Gibbs construction global neutrality of electric charge  is required.  Phase transition occurs in a broad area with smoothly increasing of energy density and baryon density.
 
 The EOS from Maxwell construction can get  smoothened   by pasta structures in the mixed phase with  sigmoid function $f_{\pm}(p)$. The parametrization of the EOS is
 \begin{equation}
 \varepsilon(p)=\varepsilon_h(p)f_{-}(p)+\varepsilon_q(p)f_{+}(p).
 \end{equation}
 Here, $ \varepsilon(p)$ is energy density in the mixed EOS while $\varepsilon_h(p)$ and $\varepsilon_q(p)$ are the energy densities in the hadronic and quark matter phases, respectively.  With this smooth EOS,   `mass-twin' phenomenon may also exist \cite{Alvarez}. In this  paper, we try to use a different interpolation method   in exploring the hybrid EOS. The hybrid EOS is derived with a  sigmoid interpolating function $f_{\pm}(\mu)$ which we will give later. The energy combination is dependent on the baryon chemical potential with a   term $ \Delta \varepsilon$ to guarantee thermodynamic consistency
 \begin{eqnarray}
\varepsilon(\mu)&=&\varepsilon^H(\mu)f_{-}(\mu)+\varepsilon^Q(\mu)f_{+}(\mu)+\Delta \varepsilon.
 \end{eqnarray}
 Here, $\varepsilon(\mu)$ is the total energy density while $\varepsilon^H(\mu)$ and $\varepsilon^Q(\mu)$ are the energy densities of hadron and quark matter respectively.
In this method, two parameters $\mu_c$ and $\Gamma$ need to be set at the beginning. The  $\mu_c$ is the central baryon chemical potential and $\Gamma$ is  half of the interpolating interval at the phase transition area. Out of the interpolating interval, the phases in the low and high chemical potential range are purely hadron and quark phase, respectively.

In Ref.\cite{ Alford2}, an additional term related to speed of sound  $c_{QM}$ in the quark matter    is added in the energy density of quark phase
 \begin{eqnarray}
\varepsilon(P)&=&\varepsilon^H(P)+\Delta \varepsilon+ c_{QM}^{-2}(P-P_{trans})
 \end{eqnarray}
 when pressure $P$ is larger than the transition pressure $P_{trans}$.
Other method to link the hadron and quark phase with first order phase transition  can be found   in Refs. \cite{Macher,Baym}, where the the combination of pressure is treated as a function that depends on the positions of the end point of hadron phase and the start point of quark phase in the $P$-$\rho$ or $P$-$\mu$ plane.

 At the quark level, we will use   our new self-consistent mean-field approximation model \cite{wangf,wangqy,zhaot,wangqw}. Based on the Nambu$-$Jona-Lasinio (NJL) model and its Fierz-transformation, a   parameter $\alpha$  is introduced to weight the contribution from Fierz-transformation Lagrangian. Here $\alpha$ in some sense is like the $\eta$ in Ref. \cite{Benic}  to count on the vector channel strengths but the way of introduction is totally different.  We can adjust $\alpha$ to get   stiffer  EOS of the quark matter and obtain quark star mass larger than 2 $M_{\odot}$. By adjusting weighting parameter $\alpha$ and the vacuum pressure of quark matter we will get the hybrid EOS and compare it with the latest astronomical observation data by calculating the mass-radius relation and tidal deformability.

 This paper is organized as follows: In Sec. II, we introduce the Walecka's nonlinear relativistic  mean-field model and our newly developed self-consistent mean-field theory of the NJL model. Then we give our   construction for hybrid EOS.  In Sec. III, We give our numerical results and analysis on the phase transition. The mass-radius relations are calculated and results are compared with the newly observed high-mass stars. Sec. IV is a short summary of our work.

\section{qurk and hadron Models  }
In construction of EOS of hybrid star, we need  on one hand  models  for quark and hadron phases  respectively, and  on the other hand a  method to connect the EOS of quark and hadron phases. In this  paper,   Walecka's relativistic mean-field (RMF) theory is used to describe the hadronic state   and  the recently developed two-flavor NJL model is used to study the quark state \cite{wangf,wangqy,zhaot,wangqw}.  The envelope of the compact object is described by the EOS of Baym-Pethick-Sutherland   \cite{BPS} and Negele-Vautherin \cite{NV}.

\subsection{The nonlinear $\sigma-\omega-\rho$ model}

 The nonlinear RMF   approach has widely being used in descriptions of nuclear matter and finite nuclei \cite{Glendenningb,Glendsigma}.
In the RMF model, the nucleon-nucleon interaction is modelled by the exchanging of  $\sigma$, $\omega$ and $\rho$ mesons. Leptons are added in a $\beta$-equilibrium system to keep chemical equilibrium and charge neutrality. In the simplest $n$-$p$-$e$ system, the effective Lagrangian can be written as:

\begin{eqnarray}
   \mathcal {L} &=&\bar\psi[i\gamma_{\mu}\partial ^\mu - M +g_\sigma \sigma -g_\omega \gamma_\mu\omega^\mu  -g_\rho\gamma_\mu  \tau_i \cdot \rho_i ^\mu ]\psi \nonumber\\
    &&+\frac{1}{2}(\partial_{\mu}\sigma\partial^\mu\sigma-m_{\sigma}^2\sigma^2) -\frac{1}{3}g_2\sigma^3-\frac{1}{4}g_3\sigma^4\nonumber\\
      & &+ \frac{1}{2}m_\omega^2\omega_\mu\omega^\mu-\frac{1}{4}\omega_{\mu\nu}\omega^{\mu\nu}+\frac{1}{4}c_3(\omega_{\mu}\omega^{\mu})^2\\
      & & +\frac{1}{2}m_ \rho  ^2  \rho_{i  \mu} \rho_i ^\mu -\frac{1}{4} \rho_ {i\mu\nu} \rho_i ^{\mu\nu} +\bar\psi_e[i\gamma_{\mu}\partial ^\mu - m_e]\psi_e \nonumber,
\end{eqnarray}
where  $\omega_{\mu\nu}$ and $\rho_{\mu\nu}$  are the  antisymmetric  tensors  of  vector  mesons
\begin{eqnarray}
  \omega_{\mu\nu}&=& \partial_\mu\omega_\nu -\partial_\nu\omega_\mu ,\\
  \rho_{i\mu\nu} &=& \partial_\mu\rho_{i\nu} -\partial_\nu\rho_{i\mu} .
\end{eqnarray}

In the mean-field approximation, the Euler-Lagrange equations reduce to  simpler forms which depend on the ground state expectations of nucleon currents. Meson fields  are  replaced by their   expectation values.
The nucleons and electrons are considered as ideal Fermi gas, and then the  requirement of charge neutrality $\rho_p=\rho_e$ gives $\mu_p=\mu_e$. Combining with the $\beta$ equilibrium  ($\mu_n=\mu_p+\mu_e$),  there is only one  free parameter (baryon chemical potential $\mu_B$ or baryon number density $\rho^H=\rho_p+\rho_n$) in solving the Euler-Lagrange equations     which gives the EOS of the hadronic system. The energy density and pressure of nuclear matter  are written as

 \begin{eqnarray}
   \epsilon^H &=& \sum_{B=n,p}\frac{1}{\pi^2}\int_0^{k_F^B}\sqrt{k^2+m^{*2}}k^2dk\nonumber\\
   &&+\frac{1}{2}m_\sigma^2\sigma^2+\frac{1}{3}g_2\sigma^3+\frac{1}{4}g_3\sigma^4 + \frac{1}{2}m_\omega^2\omega^2\\
  && +\frac{1}{4}c_3\omega^4+\frac{1}{2}m_\rho^2\rho^2 +
\frac{1}{\pi^2}\int_0^{k_F^e}\sqrt{k^2+m_e}k^2dk,\nonumber\\
   P^H &=&  \sum_{B=n,p}\frac{1}{3 \pi^2} \int_0^{k_F^B}\frac{k^4}{\sqrt{k^2+m^{*2}}}dk\nonumber\\
   && -\frac{1}{2}m_\sigma^2\sigma^2-\frac{1}{3}g_2\sigma^3-\frac{1}{4}g_3\sigma^4 +\frac{1}{2}m_\omega^2\omega^2\\
    &&+\frac{1}{4}c_3\omega^4+\frac{1}{2}m_\rho^2\rho^2
    +\frac{1}{3\pi^2}\int_0^{k_F^e}\frac{k^4}{\sqrt{k^2+m_e}}dk\nonumber,
 \end{eqnarray}
where $m^*=M-g_\sigma \sigma$ is the nucleon effective mass, $k_F$ is Fermi momentum and all meson fields ($\sigma$, $\omega$, $\rho$) denote their mean-field values.

 There are many set of  parameters in the RMF theory. We have tried the  two  typical  sets of parameters,  NL3 \cite{NL3} and TM1 \cite{TM1} .  We   found only the NL3 parameters is suitable in out construction of quark-hadron hybrid EOS.     With a saturation density $\rho_0=0.148 $ fm$^{-3}$, the NL3 parameters are $m_N=939.0$ MeV,  $m_\sigma=508.194$ MeV,  $m_\omega=782.501$ MeV,  $m_\rho=763.0$ MeV,  $g_\sigma = 10.217$, $g_ \omega=12.868 $,  $g_ \rho=4.474$,  $g_ 2=-10.431 $ fm$^{-1}$, $g_3 = -28.885$, $c_3 =0 $.  For a proton-neutron star with these parameters and without considering the EOS of outer crust, the maximum mass of neutron star is about 2.76 $M_\odot$ with radius about 12.66 km.  If one  considers hyperons in the Lagrangian, the coupling between hyperons  and mesons must  be considered   and  the EOS  will be softened.  The hyperons will not be considered in this work  since we only consider a non-strange star.  Therefore  for the quark matter  we will  also  use a two-flavor model.

\subsection{NJL model}
We investigate the deconfined quark matter within our newly developed NJL model \cite{wangqy,wangf,zhaot,wangqw}.   The NJL model is originally a model of interacting nucleons but later  used  for quarks.   It works in    the regions where perturbative QCD is not accessible \cite{Buballa,Klevansky,Kunihiro}. The simplest form of NJL Lagrangian includes only the scalar and pseudo-scalar four-quark interactions.
Its  Fierz-transformation  is just a rearrangement of fermion field operators. They can be written as

  \begin{eqnarray}
\mathcal{L}_{NJL}=\bar{\psi }(i\slashed{\partial } - m)\psi + G[\left(\bar{\psi }\psi\right)^2+\left(\bar{\psi } i \gamma ^5 \vec{\tau }\psi \right)^2],
\end{eqnarray}
 and
\begin{eqnarray}
\begin{aligned}
\mathcal{L} _{Fierz}=& \bar{\psi }(i\slashed{\partial } - m)\psi +  \frac{G} {8 N_c}[2\left(\bar{\psi } \psi \right)^2+2\left(\bar{\psi } i \gamma ^5 \vec\tau \psi \right)^2\\
& - 2\left(\bar{\psi } \vec\tau \psi \right)^2 - 2\left(\bar{\psi } i \gamma ^5 \psi \right)^2 - 4\left(\bar{\psi } \gamma^{\mu } \psi \right)^2\\
& - 4\left(\bar{\psi } \gamma^{\mu } \gamma^5\psi \right)^2 + \left(\bar{\psi } \sigma^{\mu \nu} \psi \right)^2 - \left(\bar{\psi } \sigma^{\mu \nu} \vec\tau \psi \right)^2],
\end{aligned}
\end{eqnarray}
where $m$ is the current quark mass, $G$ is the four-quark effective coupling, and $N_c$ is the number of color.
Since the two forms are equivalent under Fierz-transformation, they can be linearly combined with complex $\alpha$:

 \begin{eqnarray}
\mathcal{L} _C=\left(1-\alpha \right)\mathcal{L}_{NJL} + \alpha \mathcal{L} _{Fierz}.
\label{eq.la}\end{eqnarray}
In the mean-field approximation at finite density,
   \begin{eqnarray}\label{eq.epc1}
\left<\mathcal{L}_{NJL}\right>&=&\bar{\psi }(i\slashed{\partial } - m)\psi + 2 G \sigma_1 \bar{\psi }\psi+ \mu  {\psi^\dagger }\psi,
 \end{eqnarray}
 \begin{eqnarray}
\left<\mathcal{L} _{Fierz}\right>&=& \bar{\psi }(i\slashed{\partial } - m)\psi +  \frac{G} {2 N_c} \sigma_1 \bar{\psi } \psi +G\sigma_1^2 \nonumber \\
&&+\mu  {\psi^\dagger }\psi- \frac{G} {  N_c}\sigma_2{\psi^\dagger } \psi +\frac{G} { 2 N_c}\sigma_2^2.
 \label{eq.epc2}\end{eqnarray}
Here a term $\mu  {\psi^\dagger }\psi$ is added in both Lagrangians, with   chemical potential $\mu$ associated with  quark number density.
The two-quark condensate $\left\langle {  \bar \psi \psi} \right\rangle$ is denoted as $\sigma_1$ and $\sigma_2=\left\langle{\psi ^\dagger } \psi\right\rangle$.
From Eqs. (\ref{eq.la}$-$\ref{eq.epc2}),  the effective quark mass  and chemical potential are defined respectively, as

  \begin{eqnarray}\label{eq.muc2}
    M&=&m-2G^\prime\sigma_1,\\
 \mu_r&=&\mu- \frac{G^\prime} {  N_c}\frac{\alpha }{1-\alpha+\frac{\alpha}{4N_c}}\sigma_2. \label{eq.alpha}
  \end{eqnarray}
Here   $G^\prime$ is the four-quark effective coupling for the mixed Lagrangian Eq. (\ref{eq.la}) which has the relation with $G$
\begin{equation}\label{eq.gprime}
  G^\prime=(1-\alpha+\frac{\alpha}{4N_c})G.
\end{equation}
The new coupling $G^\prime$ needs to be recalibrated to fit the low energy experimental data.
 In the proper-time regularization, the quark condensate is given by
   \begin{eqnarray}\label{eq.gapt0}
\left\langle {  \bar \psi \psi} \right\rangle&=&-2N_c\sum_{u,d}\int \frac{d^3p}{(2\pi)^3} \frac{M}{E_p}(1-\theta(\mu-E_p))\nonumber\\
&=&-2N_c\sum_{u,d}\left( \int\frac{d^3p}{(2\pi)^3} \int_{\tau_{ UV}}^\infty d\tau  \frac{e^{-\tau E^2}}{ \sqrt{\pi \tau}} \right.  \nonumber\\
 &&\left.  -\int \frac{d^3p}{(2\pi)^3} \frac{M}{E_p}\theta(\mu_r-E_p)\right),
\end{eqnarray}
where $\tau_{UV}$ is introduced to regularize the ultraviolet divergence, and
$E_p=\sqrt{\vec{p}^2+M^2}$ defines the particle energy.
 Three parameters ($m$,   $G^{\prime}$,  and $\tau_{UV}$)  are fixed by fitting to the Gell-Mann$-$Oakes$-$Renner relation:$- 2 m \left\langle {  \bar \psi \psi} \right\rangle$ =  $(f_\pi m_\pi)^2$. Here, $f_\pi=93$ MeV,  $m_\pi= 135$ MeV, and $m = 3.5$ MeV.   The quark condensate is $\left\langle {  \bar \psi \psi} \right\rangle)^{1/3}=-282.4$ MeV. Then we have  $G^\prime=4.1433 \times 10^{-6} $MeV$^{-2}$,  and $\tau_{UV} = 955 $ MeV.
At a zero temperature, the quark number density is given by
\begin{eqnarray}
\rho_{u,d}= 2N_c \int \frac{ d^3p} {(2\pi)^3} \theta(\mu_r-E_p).
\end{eqnarray}
The quark pressure and energy density for quark matter are given by \cite{zong1,zong2}

\begin{eqnarray}
\epsilon(\mu_u,\mu_d)&=&-P(\mu_u,\mu_d)+\sum_{u,d}\mu\rho(\mu),\\
P(\mu_u,\mu_d)&=&P_0 + \sum_{u,d}\int_ 0^{\mu }d\mu \rho  (\mu ).
\end{eqnarray}
Here, $P_0$ represents the vacuum pressure density at $\mu=0$.
In many works, $P_0$ is taken as  a free parameter corresponding to the bag constant in the MIT bag model,  or   defined  as the pressure difference between   Nambu phase  and Wigner phase in the case of chiral limit.  Here we take $P_0$=$-(140 $ MeV$)^4$.
 We   include the electron to guarantee the local neutrality of electric charge with
\begin{eqnarray}
 \frac{2}{3}\rho_u-\frac{1}{3}\rho_d-\rho_e=0 .
\end{eqnarray}
Using $\mu_e$ to denote the electron-charge chemical potential and treating the electron as ideal Fermi gas, the electron density is given by
\begin{eqnarray}
\rho_e= \frac{\mu_e^3}{3\pi^2}.
\end{eqnarray}
The total baryon number density is $\rho^Q=(\rho_u+\rho_d)/3$.
The chemical potentials satisfy the weak equilibrium $d\leftrightarrow u+e+\bar\nu_e$ which gives
\begin{eqnarray}
\mu_d=\mu_u+\mu_e.
\end{eqnarray}
Then the pressure and energy density of the quark matter are
\begin{eqnarray}
\epsilon^Q&=&\epsilon(\mu_u,\mu_d)+  \frac{\mu_e^4}{4\pi^2},\\
P^Q&=&P(\mu_u,\mu_d ) +  \frac{\mu_e^4}{12\pi^2},
\end{eqnarray}
respectively.    With a fixed
 pressure of vacuum $P_0$, the stiffness of EOS increases along with $\alpha$.

\subsection{  Construction for the phase transition}

Commonly there are  two  methods to mix the quark phase and hadron phase by  requiring the chemical, thermal, and mechanical equilibrium.
  The Gibbs construction.  With a volume fraction $\chi$ of the quark phase,  it  only requires  the charge neutrality to be fulfilled globally.  In this construction, the pressure of mixed phase continuously increase with the baryon number density.
  The Maxwell construction.   It   requires the local electrical neutrality. The EOS can be obtained independently.  At the phase transition point the two different  phases have same chemical potential, temperature, and pressure with  $\mu^H=\mu^Q$,  $ P^H=P^Q$,  and $ n_q^H=n_q^Q=0$,
where $n_q$ is the local electric charge density.
 It is different from Gibbs construction that the two phases at the phase transition point have different baryon number density. It is expected for star with spherical symmetry that the pressure is a continuous function of radius and thus of baryon density. Furthermore, this construction means a very large hadron-quark surface tension and leads to  a first order phase transition between hadron and quark phases.

\begin{figure}[h]
    \includegraphics[width=1\columnwidth]{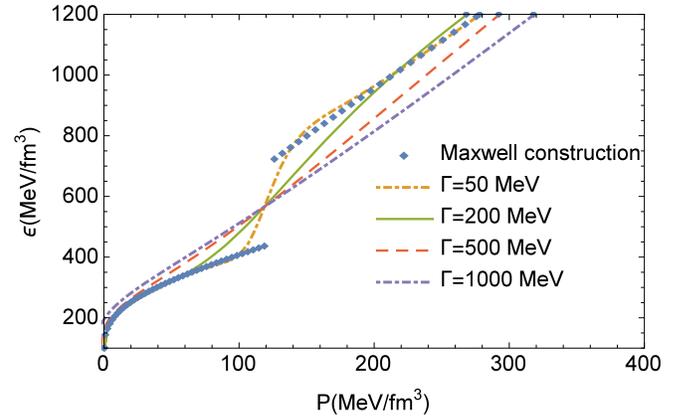}
    \caption{ EOSs at $\alpha$=0.5 for different interpolating interval $\Gamma$. The Maxwell construction of hybrid EOS   corresponds to  $\Gamma=0$. The influence of $\Gamma$ on the stiffness of EOS  at the two sides of $\mu_c$ are different.}\label{fig.gamma}
\end{figure}

 How  the nature of the phase transition  depends on the surface tension still remains unclear. Based on the analysis of the   two methods  above, to get the hybrid EOS  in this paper, we try to use an interpolating approach to connect the hadron and quark EOS. Specifically, we adopt the P-interpolation in $P-\mu$ plane. In this method, the pressure and energy density of hybrid EOS are give by
 \begin{eqnarray}
P(\mu)&=&P^H(\mu)f_{-}(\mu)+P^Q(\mu)f_{+}(\mu),\\
\varepsilon(\mu)&=&\varepsilon^H(\mu)f_{-}(\mu)+\varepsilon^Q(\mu)f_{+}(\mu)+\Delta \varepsilon, \label{eq.eos}
 \end{eqnarray}
with
 \begin{eqnarray}
f_{\pm}(\mu)&=&\frac{1}{2}\left(1\pm \texttt{tanh}\left(\frac{\mu-\mu_c}{\Gamma}\right)\right), \label{eq.sigmid}\\
  \Delta \varepsilon&=&\frac{2\mu}{\Gamma}(P^Q-P^H)(e^X+e^{-X})^{-2},\\
  X&=&\frac{\mu-\mu_c}{\Gamma}.
 \end{eqnarray}
 The function $f_{\pm}$ is a sigmoid function which has similar role as the $\chi$ in the Gibbs construction and realize a smooth EOS in the interval  of $(\mu_c-\Gamma$, $\mu_c+\Gamma)$. The additional term $ \Delta \varepsilon$ guarantees thermodynamic consistency.  The pressure $(P^H$, $P^Q)$ and  energy density $(\varepsilon^H$, $\varepsilon^Q)$ can be derived separately at the hadron phase and quark phase. But for the hybrid EOS,  two free parameters,  $\bar\mu$ and  $\Gamma$, remain  to be determined.
As showed in Figure \ref{fig.gamma}, when $\Gamma$   approaches zero, the smooth EOS  transitions to a discontinuous EOS that is a result of Maxwell construction.

\section{results and analysis}
\subsection{The hybrid EOS}

The quark  chemical potential related to baryon  chemical potentials  can be written in detail as any of the following equations:
 \begin{eqnarray}
   \mu_u &=&(\mu_n-2\mu_e)/3 ,\\
    \mu_d & = &(\mu_n+\mu_e)/3,\\
     \mu_n &=& \mu_u+2\mu_d ,
   \end{eqnarray}
where $\mu_n$ is the baryon chemical potential of neutron.

In the sigmoid function $f_{\pm}(\mu)$ of Eq. (\ref{eq.sigmid}),   $\mu_c -\Gamma$  sets  the beginning of deconfinement of   quark. The parameter $\Gamma$ affects the stiffness of the EOS. For an illustration, we plot the EOS of $\alpha=0.5$ for different $\Gamma$'s  in Figure \ref{fig.gamma}. There is a jump for $\Gamma=0$, but not for large enough $\Gamma$. As  shown  in Figure \ref{fig.gamma} for $\Gamma\geq50$ MeV, all lines of hybrid EOS  smoothly increase with pressure and have an intersection at the equilibrium pressure defined in Maxwell construction. As $\Gamma$ gets  larger than 1 GeV, the   hyperbolic tangent function approaches  zero, then the two phases will have equal weights.  Thus, it is important to set constraints on the possible values of $\Gamma$. However,  in our NJL model  only chiral transition is studied, whose relation with deconfinement transition is unclear. Some studies \cite{Fukushima1,Fukushima2} suggest that the transition starts at  roughly  1 GeV.  So, it set constraint on
$\Gamma$ with   $\Gamma$ < $\mu-1$ GeV.

\begin{figure}[h]
    \includegraphics[width=1\columnwidth]{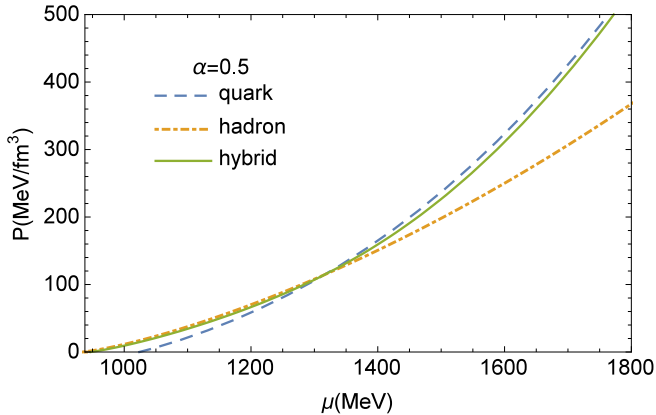}
    \caption{An illustration of the EOS with pressure-interpolation. The transition can be realized smoothly in the region of ($\mu_c -\Gamma$, $\mu_c +\Gamma$). }\label{fig.hybridpmu5}
\end{figure}

 The baryon density and pressure of the hadron matter is 2.67 $\rho_0$ and 108 MeV$/fm^3$, respectively, at  $\mu=1300$ MeV  and 2.98 $\rho_0$ and 151 MeV$/fm^3$ at  $\mu=1400$ MeV.
For the quark matter,  the baryon density at  $\mu=1300$(1400) MeV is about 4.5(5.6) $\rho_0$ for different $\alpha$ with   pressure range from  100 MeV to 400 MeV.  An illustration of the interpolation in the $P-\mu$ plane is presented in Figure \ref{fig.hybridpmu5}.

\begin{figure}[h]
    \includegraphics[width=1\columnwidth]{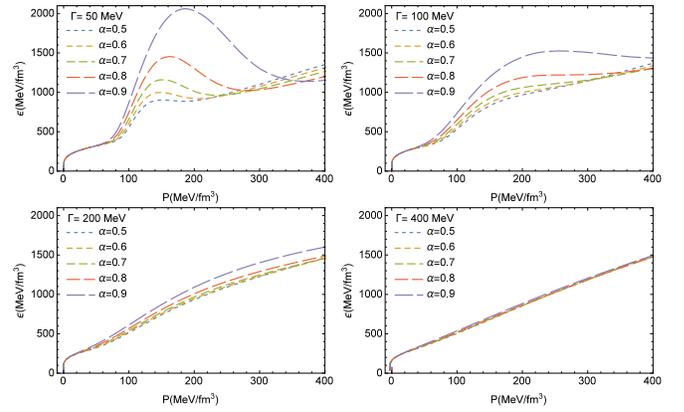}
    \caption{ Hybrid EOS for different interpolation width $\Gamma$ at $\mu_c=1300$ MeV.  The results of different $\alpha$ begin to differentiate when $\Gamma$ becomes smaller.   }\label{fig.hybrideos}
\end{figure}

  Plots of the energy density as function of pressure
 are presented in Figure \ref{fig.hybrideos}. The interpolation width $\Gamma$ has significant influences on the stiffness of hybrid EOS at the two sides of $\bar\mu$. At large $\Gamma$, the lines monotonically increase. When  $\Gamma$ is relatively small,   the lines become non-monotonically increasing around  $P=150$ MeV$/fm^3$ which is the pressure of hadron matter at $\mu=1300$ MeV.
The weighting parameter $\alpha$ has significant impact on the stiffness of hybrid EOS. The stiffness of quark EOS increases with $\alpha$, but this only happens for small $\Gamma$ at high pressure in the hybrid EOS. This confirms that the hadron EOS is  dominant  at low density.

\begin{figure}[h]
    \includegraphics[width=1\columnwidth]{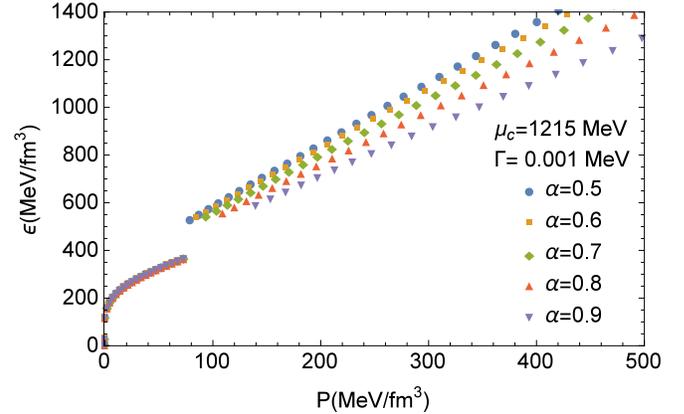}
    \caption{ Hybrid EOS for different $\alpha$  at $\mu_c=1215$ MeV. In this case, the EOS from the interpolation method is corresponding to the EOS from  Maxwell construction.   }\label{fig.eosmax}
\end{figure}

 To produce the EOS of Maxwell construction, it is not just to set $\Gamma=0$ with arbitrary $\mu_c$. We must find the intersection of the quark and hadron EOS in the $P-\mu$ plane. The transition chemical potential is 1215 MeV for $\alpha=0.5$. We plot the EOS that corresponds to the one from Maxwell construction in Figure \ref{fig.eosmax}.  For $\alpha$ larger than 0.5 at $\mu=1215$ MeV, both energy density and pressure are discontinuous. 

\subsection{Mass-radius relations}

  We use  the  TOV equations 
(in units $G=c=1$ ):
\begin{eqnarray}
 \frac {dP\left(r \right)} {dr} &=&-\frac {\left(\epsilon +P \right)\left(M + 4\pi r^3 P \right)} {r \left(r-2M \right)},\\
\frac {dM\left(r \right)} {dr} &=&4\pi r^2 \epsilon,
 \end{eqnarray}
to investigate the mass-radius relation.

\begin{figure}[h]
    \includegraphics[width=1\columnwidth]{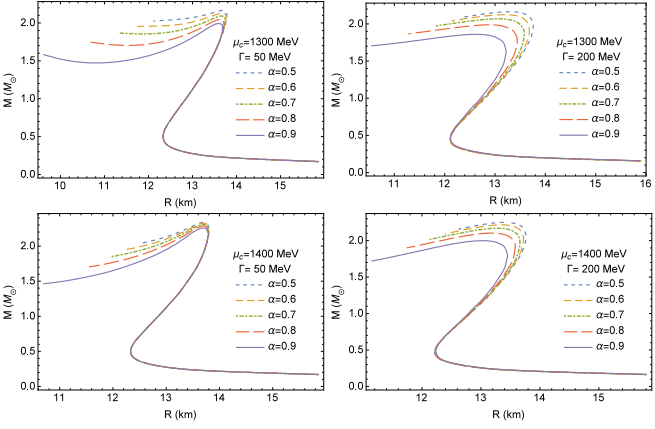}
    \caption{ The mass-radius relation for different $\alpha$, $\Gamma$, and $\mu_c$.  }\label{fig.mr}
\end{figure}

 The mass-radius relations are showed in Figure \ref{fig.mr} which has several features.
First of all, when the $\Gamma$ and  $\mu_c$ are fixed, the difference of hybrid EOS mainly lies in the difference of $\alpha$, and then affects the mass radius relation.
The maximum mass of hybrid stars increases with increasing $\alpha$, which is different with that of pure quark stars. As shown in the upper-left picture in Figure \ref{fig.hybrideos}, the hybrid EOS becomes softer with increasing $\alpha$ in the middle region of pressure.
Second,   we use the harder EOS of hadron matter that neutron star can have maximum mass about 2.7 $M_\odot$. In the Figure \ref{fig.mr}, the maximum mass of the hybrid star appears below 2.5 $M_\odot$.
The  curves almost coincide with each other in low mass regions, but separate with the change of $\alpha$ in high mass regions.
This shows that for hybrid stars from our model,   hadron matter dominates in low mass stars, while for high mass stars, quark matter has a great influence on the star mass.
Third, when $\Gamma$ is fixed, the maximum mass of the hybrid star increases with increasing chemical potential $\mu_c$.
When   $\mu_c$ is fixed, quark matter and hadron matter can be mixed in a larger chemical potential region with increasing $\Gamma$. This is reflected in the Figure \ref{fig.mr} where the   curves begin to separate at a smaller mass.
Lastly, in the upper-left picture of Figure \ref{fig.mr}, it seems that mass twin phenomenon appears for larger $\alpha$ when extending the curves.  But the energy density and pressure of the hybrid EOS given by our model is not high enough.  We need to do further analysis on this point.

 In Figure \ref{fig.mr},   the radius for a 1.4 M$_\odot$ star is  less than the constraints   $R \leq 13.76~$km or $R \leq 13.6~$km \cite{Fattoyev,Annala}  and  larger than $10.7~$km  \cite{Bauswein2019}. Although these radius constraints are model-dependent and may not suitable for  hybrid stars, we  still present here as  a comparison.

\subsection{ Tidal deformability}

 In the early works  of Ref. \cite{Margali,Abbott},  restrictions on the tidal deformability  for a $1.4~  M_{\odot} $ is  less than 800~(1400)~ for low~(high)~--spin prior case .
 The most recently analysis  in analysis on the binary neutron star merger GW170817 have found tighter constraints on the component mass to lie between  1.00  and 1.89 $M_\odot$
with $\tilde\Lambda$ in (0, 630)  when allowing for large component spins and
 on the component masses to lie between   1.16  and 1.60  $M_\odot$ with $\tilde\Lambda = $ $300^{+420}_{-230}$  when the spins are restricted to be within the range observed in Galactic binary neutron stars \cite{Abbott2019}.

 The $\tilde\Lambda$ is a mass-weighted linear combination of the two tidal
parameters $ \Lambda_1$ and $ \Lambda_2$. With $M_1$  and $M_2$  the corresponding gravitational masses, the $\tilde\Lambda$ is defined as
 \begin{equation}\label{eq.tildelambda}
 \tilde\Lambda=\frac{16}{13}\frac{(M_1+12M_2)M_1^4\Lambda_1+(M_2+12M_1)M_2^4\Lambda_2}{(M_1+M_2)^5}.
 \end{equation}
In deducing the  $\tilde\Lambda$ in the low-spin case, the component mass have being constrained to  $M_1$ $\in$ (1.16,1.36) $M_\odot$ and $M_2$ $\in$ (1.36,1.60) $M_\odot$ with total mass $2.73^{+0.04}_{-0.01}$$M_\odot$ and chirp mass $1.186^{+0.001}_{-0.001}$$M_\odot$. The chirp mass is defined as
 \begin{equation}\label{eq.chirpmass}
  M_{chirp}=\frac{(M_1M_2)^{3/5}}{(M_1 + M_2)^{1/5}}
 \end{equation}

  In the calculation of tidal deformability $\Lambda$,  one way is to use a  universal relation only between $\Lambda$  and compactness $C= M/R$ with $M$ and $R$ being the star mass and radius respectively \cite{Carson,Chen}. But here, we calculate  $\Lambda$ through the compactness and the Love number.
The tidal deformability is related to the $l=2$ dimensionless tidal
Love number $k_2$ through 
\begin{equation}\label{eq.klam}
  k_2=\frac{3}{2}\Lambda\left(\frac{M}{R}\right)^5
\end{equation}
in units $G=c=1$.
The $l=2$ tidal Love number $k_2$ for the internal solution is given by  \cite{Damour}
\begin{eqnarray}
k_2&=&\frac{8}{5} C ^5   (1 - 2 C)^2 [2 + 2 C (y - 1) - y] \nonumber\\
&& \times \{2 C [6- 3 y+ 3 C (5 y - 8)]\nonumber\\
&& + 4 C^3 [13 - 11 y + C (3 y - 2)+ 2 C^2 (1 + y)] \nonumber\\
&&+    3 (1 - 2 C)^2 [2 - y+ 2 C (y - 1)] ln(1 - 2 C)\}^{-1},\nonumber\\
\end{eqnarray}
in matching the interior and exterior solutions across
the star surface.
 Here, $y$ is related to the metric variable $H$ and surface energy density $\epsilon_0$

\begin{equation} \label{eq.y}
y=\frac{R\beta(R)}{H(R)}-\frac{4\pi R^3 \epsilon_0}{M}.
\end{equation}
 For some neutron star model the surface energy density is zero. But in our NJL model with negative vacuum pressure, the surface energy $\epsilon_0$ is nonzero.   So the last term can not be neglected.

 The  metric variable $H$ related to the   EOS can be obtained by integrating two differential equations
 \begin{eqnarray} \label{eq.h}
 &&\frac{dH(r)}{dr}=\beta, \\
 \frac{d\beta(r)}{dr} &=&2 gH\{-2 \pi[5\epsilon + 9 P + f (\epsilon +  P)]\nonumber\\
   && + \frac{3}{r^2} +  2 g (\frac{M}{r^2} + 4 \pi r P)^2\}+\nonumber\\
 &&  2g \frac{\beta}{r} [-1 + \frac{M}{r} + 2 \pi r^2 (\epsilon -  P)],
  \end{eqnarray}
where $g=(1-2M/r)^{-1}$ and $f=d\epsilon /dP $. The iteration start from the center at $r=0$ via expansions $H(r) =a_0 r^2 $ and $\beta(r)=2a_0r$ with constant $a_0$.  As can be seen from Eq. (\ref{eq.y}), we only concern the ratio $\beta/H$. So $a_0$ can be arbitrarily chosen   in numerical calculation.

\begin{table}[h]
  \caption{ Results of tidal deformability  with mass constraints from Ref. \cite{Abbott2019} in the low-spin case.  Here, the parameters in the interpolation are $\Gamma=50$ MeV and $\mu_c=1300$ MeV.    }
 \label{Tab.tidalchirp}
 \begin{center}
 \begin{tabular}{  c c c c  c}
    \hline   \hline

  $M_1$$(M_\odot)$ & $M_2$$(M_\odot) $  & $\Lambda_{1 }$& $\Lambda_{2}$ & $ \tilde\Lambda$\\
 \hline 
  1.160 &1.609 & 1952.10 &371.394 &839.237 \\
  1.326 & 1.400 &1027.41 & 768.748& 888.738\\
   1.362 &1.363 & 887.325 &883.712 & 885.517\\
  \hline
   \hline
 \end{tabular}
\end{center}

\end{table}

 In Table \ref{Tab.tidalchirp}, we show the calculated  $\Lambda$ and $\tilde\Lambda$ for masses satisfying  the mass constraints given above, i.e., $M_1$ $\in$ (1.16,1.36) $M_\odot$ and $M_2$ $\in$ (1.36,1.60) $M_\odot$ with total mass $2.73^{+0.04}_{-0.01}$$M_\odot$ and chirp mass $1.186^{+0.001}_{-0.001}$$M_\odot$.
To include the boundary and mass of 1.4 $M_\odot$, we design three sets of mass as shown in the table.
 First, we choose a narrow width $\Gamma=50$  MeV with $\mu_c=1300 $ MeV.  As shown in Figure \ref{fig.mr}, for   the mass between (1.16,1.60)
 the curve does not show a dependence on $\alpha$, so the hybrid stars are dominated by hadron matter in this mass range. The calculated results of $\tilde \Lambda$ in Table \ref{Tab.tidalchirp} are  out of the range $300^{+420}_{-230}$ in the low-spin case. 

 Then we take a larger value of $\Gamma$.  When $\Gamma$ is increased, quark matter and hadron matter are considered to mix in a larger region in the $P-\mu$ plane. We can find in Table \ref{Tab.tidalchirp2} that the calculated tidal deformabilities are within $300^{+420}_{-230}$ from   analysis on the binary neutron star merger GW170817.
In Table \ref{Tab.tidalchirp3}, we also give results when $\Gamma$ is increased further. Compared with the results in Table \ref{Tab.tidalchirp2}, the tidal deformabilities $\tilde\Lambda$ are obviously reduced but still within the allowed range. The results  are more  close  to the center value.

\begin{table}[h]
  \caption{ Results of tidal deformability with  $\Gamma=300$ MeV and $\mu_c=1300$ MeV for different $\alpha$.
   Here, the labels a, b, and c  indicate   the three sets of component mass as listed in Table \ref{Tab.tidalchirp}.
    }
 \label{Tab.tidalchirp2}
 \begin{center}
 \begin{tabular}{  c c c  c}
    \hline   \hline

$\alpha$    & $\tilde\Lambda_{a }$& $\tilde\Lambda_{b}$ & $ \tilde\Lambda_{c }$\\
 \hline 
0.5 &    482.991 & 515.836 &517.420  \\
 0.6 &   475.476 & 518.318& 520.031\\
 0.7  & 474.121 &507.478 & 506.659\\
  0.8   &470.32 & 503.15& 506.235\\
 0.9  & 459.558 &490.097 & 492.191\\

  \hline
   \hline
 \end{tabular}
\end{center}

\end{table}

\begin{table}[h]
 \caption{ Results of tidal deformability at $\alpha=0.5$ with  interpolation parameters   $\Gamma=500$ MeV and $\mu_c=1300$ MeV.    }
 \label{Tab.tidalchirp3}
 \begin{center}
 \begin{tabular}{  c c c c  c}
    \hline   \hline

  $M_1$$(M_\odot)$ & $M_2$$(M_\odot) $  & $\Lambda_{1 }$& $\Lambda_{2}$ & $ \tilde\Lambda$\\
 \hline 
  1.160 &1.609 & 753.687 &151.865 &330.689 \\
  1.326 & 1.400 &406.400 & 304.145& 351.581\\
   1.362 &1.363 & 351.925 &354.034 & 352.981\\

  \hline
   \hline
 \end{tabular}
\end{center}

\end{table}

\section{summary }

 Investigation of QCD matter transition at low temperature and high baryon chemical potential relies heavily on effective models for lacking data from experiment. Among it, a construction connecting the hadron and quark phases is indispensable.   There are two commonly used criteria for the phase transition in construct a hybrid EOS, i.e., the Gibbs construction and the Maxwell construction.  During the hadron-quark transition,  the baryon number density changes continuously by using the Gibbs construction,  but not continuously  with the Maxwell construction. Since the transition still remains uncertain  experimentally,
we take   $P$-interpolation on the $P$-$\mu$ plane to get the EOS of hybrid stars. This method, depending on the central baryon chemical potential of interpolating area $\bar\mu$ and half of the interpolating interval $\Gamma$, gives a   smooth EOS.

We use Walecka's RMF theory to describe the nuclear phase and use our recently developed NJL model to describe the quark phase.  We have tried two sets of parameters (TM1 and NL3) in the RMF theory and found only the NL3 parameter set can be used in our analysis. At the quark level, we have two parameters in adjusting the EOS.  It is the vacuum pressure $P_0$ and the parameter $\alpha$ that weights the contribution from the vector channel in the Fierz-transformed Lagrangian.

 We have studied the dependence of hybrid EOS and mass-radius relation on $\bar\mu$ and $\Gamma$. The results show that the stiffness of the EOS and maximum mass given by the hybrid EOS are sensitive   to   $\bar\mu$ and $\Gamma$. In this paper, we use the possible deconfinement chemical potential and baryon chemical potential equilibrium to fix $\bar\mu$ and $\Gamma$.    Parameters $\bar{\mu}$ and $\Gamma$ as a function of $\alpha$ give the  area of mixed phases.  The central baryon density is about $3$ $ \rho_0$ for hadron matter and about $4.5$ $\rho_0$ for quark matter, where $\rho_0$ is the saturation density of nuclear matter.

The stiffness of hybrid EOS   increases with $\alpha$. By adjusting $\alpha$, maximum mass of hybrid stars can be larger than masses from PSR J$1614$-$2230$  $ (M=1.928\pm 0.017~ M_{\odot})$  \cite{Fonseca} and PSR J$0348$+$0432$ $ (M=2.01\pm 0.04 ~M_{\odot})$ \cite{Antoniadis}  and the recent observation of MSP J0740+6620 with star mass $2.14^{+0.10}_{-0.09}$ within the 68.3  credibility interval  \cite{Cromartie}.
The calculated  radii of a 1.4-solar-mass star for two different quark vacuum pressures are less than  $R \leq 13.76~$ km  or $R \leq 13.6~$ km from Refs. \cite{Fattoyev,Annala} which are   model-dependent constrains.    The lower limit  $10.7$ km  of  a 1.6-solar-mass neutron star is  also   satisfied.
 In a recent analysis on the binary neutron star merger GW170817, the tidal deformability is   $\Lambda = $ $300^{+420}_{-230}$  for the component masses to lie between   1.16 $M_\odot$  and 1.60  $M_\odot$ when the spins are restricted \cite{Abbott2019}.

 When the interpolation area of the two phases is narrow,  mass-twin phenomenon that have been found in the Maxwell construction seems to appear in the mass-radius plot.  But under the current parameter choices, the core energy of stars can not reach the required high  value. Moreover, in the mass range of (1.16,1.36) $M_\odot$, the stars are mainly of hadron matter. The calculated tidal deformabilities $\tilde \Lambda$ do not satisfy the  results in Ref. \cite{Abbott2019}.  When $\Gamma$ is increased, we have found that the calculated $\tilde \Lambda$  are in the range  $\Lambda = $ $300^{+420}_{-230}$ of low spin case.  The results are better when $\Gamma$  increases further.
 So,  EOS of pure hadron phase   or   hybrid phase  with Maxwell construction can be excluded
by the observation of tidal deformability from GW170817,  but hybrid phase with a crossover transition
 may is still suggested to be effective to describe
the  phase in hybrid star.


\acknowledgments
This work is supported in part by the National Natural Science Foundation of China (under Grants No. 11475085,  No. 11535005, No. 11690030, No.11873030, and No. 11905104) and the National Major state Basic Research and Development of China (Grant No. 2016YFE0129300).

\end{document}